\documentclass[aps,prd,floatfix,nofootinbib,superscriptaddress,showpacs,twocolumn]{revtex4-1}

\usepackage{amssymb}
\usepackage[intlimits]{amsmath}
\usepackage{amsfonts}
\usepackage{dsfont}
\usepackage{subfigure}
\usepackage[colorlinks=true,linkcolor=blue,citecolor=blue,urlcolor=blue]{hyperref}
\usepackage{graphicx}
\usepackage[usenames]{color}
\usepackage{natbib}
\usepackage{multirow}
\usepackage{verbatim}
\usepackage{slashed}
\usepackage{nicefrac}
\usepackage[latin1]{inputenc} 

\def\F{\ensuremath{\mathcal{F}}}
\def\q{\text{q}}

\newcommand{\Tr}{\,\mathrm{Tr}\,}

\definecolor{violet}{RGB}{111,0,255}
\definecolor{webgreen}{rgb}{0,0.75,0}
\definecolor{webred}{rgb}{0.75,0,0}
\definecolor{webblue}{rgb}{0,0,0.75}
\definecolor{darkblue}{rgb}{0,0,0.6}
\definecolor{darkgreen}{rgb}{0,0.5,0.5}
\definecolor{darkpurple}{rgb}{0.5,0,0.5}
\definecolor{darkorange}{rgb}{1,0.5,0}
\definecolor{darkgrey}{rgb}{0.4,0.4,0.4}
\definecolor{lgray}{rgb}{0.95,0.95,0.95}
\definecolor{lgreen}{rgb}{0.95,1.00,0.90}
\definecolor{lred}{rgb}{1.00,0.90,0.80}
\definecolor{lblue}{rgb}{0.2,0.35,1.00}
\definecolor{shadecolor}{rgb}{1.00,0.92,0.82}

\definecolor{violet}{RGB}{111,0,255}
\definecolor{dgreen}{rgb}{0.1,0.50,0.1}

\begin{document}

\title{Single pseudoscalar meson pole and pion box contributions to the anomalous magnetic moment of the muon}

\author{Gernot Eichmann}
\email[e-mail: ]{Gernot.Eichmann@tecnico.ulisboa.pt}
\affiliation{CFTP, Instituto Superior T\'ecnico, Universidade de Lisboa, 1049-001 Lisboa, Portugal}
\author{Christian S. Fischer}
\email[e-mail: ]{Christian.Fischer@physik.uni-giessen.de}
\affiliation{Institut f\"ur Theoretische Physik, Justus-Liebig Universit\"at Gie{\ss}en, 35392 Gie{\ss}en, Germany}
\author{Esther Weil}
\email[e-mail: ]{Esther.D.Weil@physik.uni-giessen.de}
\affiliation{Institut f\"ur Theoretische Physik, Justus-Liebig Universit\"at Gie{\ss}en, 35392 Gie{\ss}en, Germany}
\author{Richard Williams}
\email[e-mail: ]{Richard.Williams@theo.physik.uni-giessen.de}
\affiliation{Institut f\"ur Theoretische Physik, Justus-Liebig Universit\"at Gie{\ss}en, 35392 Gie{\ss}en, Germany}

\begin{abstract}
We present results for single pseudoscalar meson pole contributions and pion box contributions
to the hadronic light-by-light (LBL) correction of the muon's anomalous magnetic moment. We follow
the recently developed dispersive approach to LBL, where these contributions are evaluated with
intermediate mesons on-shell. However, the space-like electromagnetic and transition form factors
are not determined from analytic continuation of time-like data, but directly calculated
within the functional approach to QCD using Dyson-Schwinger and Bethe-Salpeter equations. This strategy
allows for a systematic comparison with a strictly dispersive treatment and also with recent results
from lattice QCD. Within error bars, we obtain excellent agreement for the pion electromagnetic and
transition form factor and the resulting contributions to LBL. In addition, we present results
for the $\eta$ and $\eta'$ pole contributions and discuss the dynamical effects in the $\eta-\eta'$
mixing due to the strange quarks. Our result for the total pseudoscalar pole contributions is
$a_\mu^{\text{PS-pole}}   = 91.6 \,(1.9)  \times 10^{-11}$ and for the pion-box contribution we obtain $a_\mu^{\pi-\text{box}}   = -16.3 \,(2)(4)  \times 10^{-11}$.
\end{abstract}

\maketitle


%
%
%
%
%
\section{Introduction}\label{sec:introduction}
    The anomalous magnetic moment $a_\mu = \frac{1}{2}(g-2)_\mu$ of the muon is currently under intense scrutiny from both theory and experiment. With a persistent discrepancy of about $3$--$4$ standard deviations between the theoretical Standard Model (SM) predictions and experimental determinations~\cite{Blum:2013xva}, $a_\mu$ is considered a potential candidate for the observation of physics beyond the SM. In order to identify such contributions, both theory and experiment need to improve their precision beyond the $0.54$ parts per million level that has been achieved by E821 at Brookhaven~\cite{Bennett:2006fi,Roberts:2010cj}. Two new experiments at Fermilab~\cite{Venanzoni:2014ixa} and J-PARC~\cite{Otani:2015jra} are under way, aiming to reduce the experimental error by a factor of four.

\begin{figure}[t]
      \begin{center}
        \includegraphics[scale = 1]{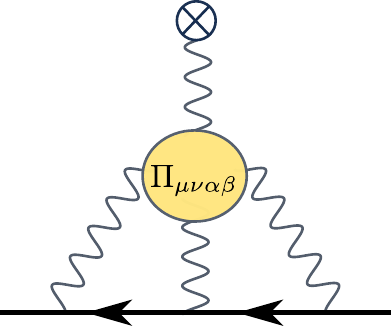}
      \end{center}
      \caption{The light-by-light scattering contribution to $a_\mu$.
      The main ingredient is the hadronic photon four-point function $\Pi_{\mu\nu\alpha\beta}$.}
      \label{fig:LBLContribution}
    \end{figure}

    However, the error budget of the theoretical SM prediction is dominated by hadronic contributions that probe non-perturbative QCD and at present mask any potential signals of new physics. The most relevant of these are hadronic vacuum polarisation (HVP) and light-by-light (LBL) scattering effects; the latter of which are the focus of this work and are shown diagrammatically in Fig.~\ref{fig:LBLContribution}.
    While the currently accepted estimate on hadronic LBL stems from a combination of calculations based on low-energy effective models~\cite{Prades:2009tw}, see \cite{Nyffeler:2017ohp} for a recent overview, there are great efforts both from lattice QCD~\cite{Green:2015sra,Blum:2015gfa,Green:2015mva,Asmussen:2016lse,Asmussen:2017bup,Asmussen:2018ovy,Blum:2017cer,Blum:2016lnc,Jin:2016rmu,Gerardin:2017ryf,Meyer:2018til} as well as dispersion theory~\cite{Colangelo:2014dfa,Colangelo:2014pva,Colangelo:2015ama,Pauk:2014rfa,Nyffeler:2016gnb,Danilkin:2016hnh,Colangelo:2017fiz,Colangelo:2017qdm,Hoferichter:2018dmo,Hoferichter:2018kwz} to improve this estimate.

    \begin{figure}[b]
      \begin{center}
         \includegraphics[scale = 1]{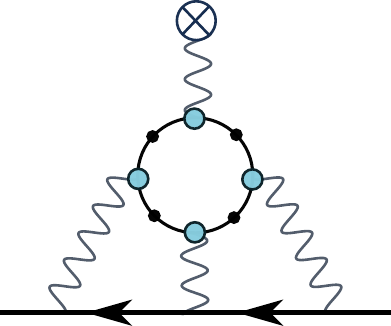} \hfill
         \includegraphics[scale = 1]{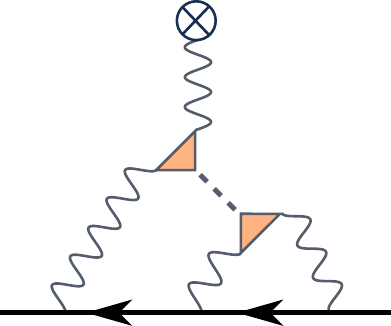}
      \end{center}
      \caption{\textit{Left:} The quark loop contribution to $a_\mu$ (without permutations of the photon legs). The quark propagators
      and quark-photon vertices are fully dressed. \textit{Right:} The meson-exchange part of the LBL contribution to $a_\mu$ (without permutations of the photon legs).}
      \label{fig:lbl_contributions}
    \end{figure}

    Within the functional approach via Dyson-Schwinger and Bethe-Salpeter equations (DSEs and BSEs),
    meson exchange contributions to LBL as well as an (incomplete) determination of quark-loop effects
    (see Fig.~\ref{fig:lbl_contributions}) have been presented and discussed in Refs.~\cite{Fischer:2010iz,Goecke:2010if,Goecke:2012qm}.
    In the same framework, the dispersive results for hadronic vacuum polarisation have been reproduced on the level of 2-3 percent~\cite{Goecke:2011pe}.

    A principal challenge for the functional approach is to provide a reliable error estimate. In all practical calculations the tower of DSEs must be truncated, and it is extremely hard to quantify the systematic error of neglected contributions. Within the class of rainbow-ladder truncations employed thus far for $a_\mu$, insight can be gained only through comparison with both experimental results and other approaches whose error estimates are well-defined.

    Subsequently, when a given truncation scheme is known to perform well for certain observables, it can be expected to perform equally well for related ones. Fortunately, the rainbow-ladder scheme used in the context of $a_\mu$ passes this test. As summarised e.g. in~\cite{Eichmann:2016yit}, it does extremely well in the pseudoscalar meson sector and very reasonably in the vector meson channels. This includes observables such as masses, form factors, charge radii and transition form factors which are all highly relevant for the calculation of $a_\mu$. Given this quality, it is plausible to make use of functional methods as a complementary tool to lattice QCD and dispersive approaches.

    In this work we use previously obtained results for the pion electromagnetic form factor (EMFF) and
    the pion two-photon transition form factor (TFF) in the DSE/BSE framework to determine the dispersive pion box and pion pole contributions to hadronic LBL. Based on the excellent agreement with recent data driven dispersive results, we then derive predictions for the $\eta$ and $\eta'$ meson pole contributions and discuss the impact of the strange quark dynamics. In the following we briefly summarise the technical elements of our calculation  followed by a discussion of the results.
    We use a Euclidean notation throughout this work; see e.g. Appendix A of Ref.~\cite{Eichmann:2016yit} for conventions.

\section{Anomalous Magnetic moment}
    To obtain the LBL contribution to the muon anomalous magnetic moment $a^\mathrm{LBL}_\mu$,
    one must consider its contribution to the muon-photon vertex shown in Fig.~\ref{fig:LBLContribution}. On the muon mass-shell this vertex can be decomposed as
    \begin{align}
       \parbox{1cm}{\includegraphics[scale = 0.5]{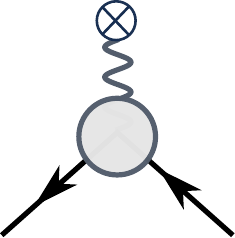}}\quad
       &=\bar{u}(p')
       \left[F_1(Q^2)\gamma^\alpha - F_2(Q^2)\,\frac{\sigma^{\alpha\beta} Q^\beta}{2 m_\mu}\right]u(p),
      \label{eqn:MuonPhotonVertexDecomposition}
    \end{align}
    where $p$ and $p'$ are the muon momenta, $Q$ is the photon
    momentum and $\sigma^{\alpha\beta}=-\frac{i}{2}[\gamma^\alpha,\gamma^\beta]$. The
    anomalous magnetic moment is defined as
    \begin{align}
      a_\mu = \frac{g-2}{2}=F_2(0),
      \label{eqn:DefOfAnomaly}
    \end{align}
    which is obtained from Eq.~(\ref{eqn:MuonPhotonVertexDecomposition}) in the limit of vanishing photon momentum $Q^2$.

    In order to extract $a_\mu$ we use the technique advocated in Ref.~\cite{Aldins:1970id}, see also Ref.~\cite{Goecke:2010if}
    for details. We then obtain
    \begin{equation}\label{amu}
      a_\mu = \frac{im_\mu}{12}\lim_{Q\to 0} \,\text{Tr}\left\{ [\gamma^\rho, \gamma^\sigma]\,\Gamma^{\rho\sigma}\right\}
    \end{equation}
    with the muon-photon vertex
    \begin{equation}\label{photon_four_point}
    \begin{split}
       \Gamma^{\rho\sigma} =& \int_{q_1}\!\!\int_{q_3} \Lambda_+(p')\,\gamma^\mu \, S(k_1) \, \gamma^\nu \, S(k_2) \, \gamma^\lambda \,\Lambda_+(p)   \\
       & \times D^{\mu\mu'}(q_1)\,D^{\nu\nu'}(q_2) \,D^{\lambda\lambda'}(q_3)  \\
       & \times  \frac{\partial}{\partial Q^\rho}\Pi^{\mu'\nu'\lambda'\sigma}(q_1, q_2, q_3, Q)\,,
    \end{split}
    \end{equation}
    including a derivative with respect to the momentum $Q$ of the external photon. Here, $m_\mu$ denotes the
    muon's mass, $S$ its propagator and $\Lambda_+$ its positive-energy projector,
    $D^{\lambda\lambda'}$ is the photon propagator and $\Pi^{\mu'\nu'\lambda'\sigma}$
    the photon four-point function.
    We abbreviated the momentum integration in four dimensions by $\int_{q} := \int d^4q/(2\pi)^4$.

    \begin{figure}[b]
      \begin{center}
         \includegraphics[scale = 0.8]{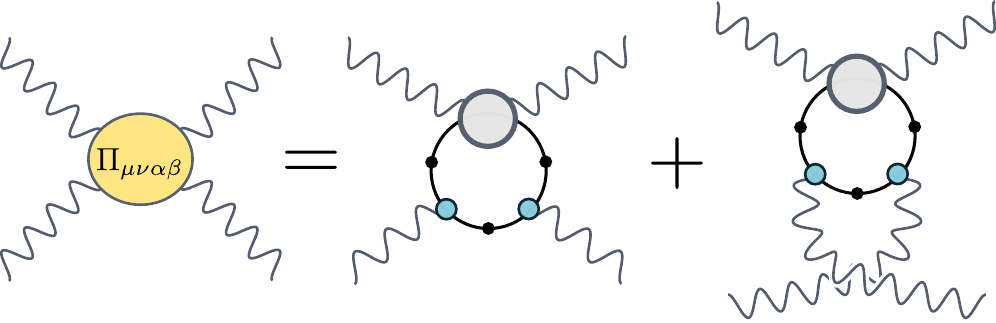}
      \end{center}
      \caption{Representation of the photon four-point function in terms of the quark Compton vertex 
      (see App. B in Ref.~\cite{Goecke:2012qm} for a derivation).
      The quark-photon four-point function in the figure is defined as $\widetilde\Gamma - \tfrac{1}{2} \Gamma_B$,
      where $\widetilde\Gamma$ is the full quark Compton vertex defined in Ref.~\cite{Eichmann:2012mp} and $\Gamma_B$ are the Born terms.
      Terms beyond
      the rainbow-ladder truncation used in this work are already omitted, see~\cite{Eichmann:2012mp} for
      the full expansion.}
      \label{fig:compton}
    \end{figure}

\subsection{Single meson pole contributions}\label{sec:mesonpole}

    The photon four-point function in
    Eq.~(\ref{photon_four_point}) can be approximated by expanding it in terms of various hadronic contributions.
    Working directly in the space-like momentum domain that is characteristic for the LBL integral,
    an expansion in terms of quark and gluon degrees of freedom and a subsequent
    resummation into hadronic degrees of freedom has been discussed in detail in Ref.~\cite{Goecke:2010if}. It agrees
    with standard treatments within effective models, see the review \cite{Jegerlehner:2009ry} and references therein.
    The leading terms in this expansion are the quark-loop and meson exchange diagrams shown in Fig.~\ref{fig:lbl_contributions}.
    In such a framework the exchanged mesons have to be considered off-shell, which requires a non-unique (but IR or UV-constrained)
    prescription for the off-shell meson propagators and form factors. While in principle the expansion in terms of
    quark and gluon degrees of freedom can be treated in a unique and well-defined manner using the representation
    of Fig.~\ref{fig:compton} and treating the quark-Compton vertex along the lines of
    Refs.~\cite{Eichmann:2012mp,Eichmann:2018ytt}, in practice this is currently not feasible due to unsolved problems with
    transversality and analyticity in the gauge dependent basis elements of the photon four-point function, see
    \cite{Eichmann:2015nra} for details.

    \begin{figure*}[t]
      \begin{center}
         \includegraphics[scale = 0.82]{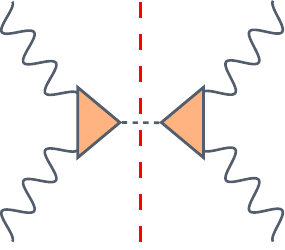} \hspace*{2cm}
         \includegraphics[scale = 0.82]{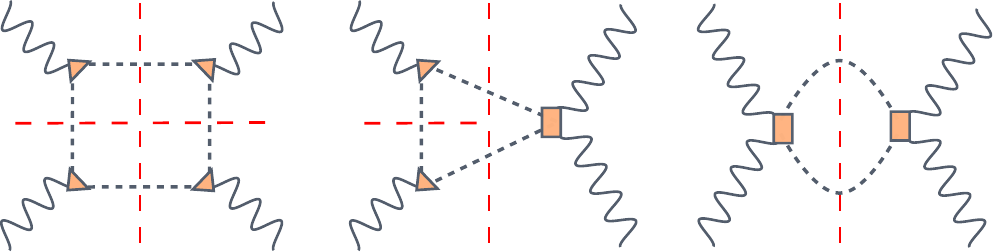}
      \end{center}
      \caption{Dispersive pseudoscalar meson-pole (left) and pion-loop (right) contributions to the photon four-point function.}
      \label{fig:dispersive}
    \end{figure*}

    In contrast, the dispersive approach offers a unique expansion in terms of diagrams involving one or more intermediate
    mesons \cite{Colangelo:2014pva,Colangelo:2014dfa,Pauk:2014rfa,Colangelo:2015ama}. The corresponding
    diagrams for the (leading) pseudoscalar meson pole contributions and the pion box diagram are shown in Fig.~\ref{fig:dispersive}.
    This expansion is genuinely different than the one of Fig.~\ref{fig:lbl_contributions}. Although superficially the meson
    pole diagram looks very similar to the corresponding diagram in Fig.~\ref{fig:lbl_contributions}, they are not the same.
    In the dispersive approach the exchanged meson and the two TFFs are evaluated as on-shell quantities,
    in contrast to the off-shell nature of the 'resummed mesons' considered above. For pseudoscalar
    mesons the meson pole diagram is given by the permuted sum of two meson TFFs coupled with
    an appropriate propagator,
    \begin{align}\label{mesonpole}
      \Pi^{\mu\nu\alpha\beta}(q_1,q_2,q_3,q_4) =&&\\
      \sum_M \mathcal{F}_M^{\mu\nu}(q_1,q_2)
      D_{M}&(q_1+q_2)
      \mathcal{F}_M^{\alpha\beta}(q_3,q_4) + \mathrm{(perms)}.\nonumber
    \end{align}
    Here, $\mathcal{F}_M^{\mu\nu}(q_1,q_2)$ is the two-photon TFF for meson $M$ with
    $D_M(q_1+q_2)$ its free propagator. Together with Eqs.~(\ref{amu}),
    (\ref{photon_four_point}) this general expression can be drastically simplified using projection and integration techniques
    in terms of Gegenbauer polynomials. As has been shown in \cite{Colangelo:2015ama}, the special case of the pion pole
    contribution eventually reduces to a simple well-known formula that has been developed earlier \cite{Knecht:2001qf}
    already in the context of effective models.

    In previous calculations in the functional DSE framework \cite{Fischer:2010iz,Goecke:2010if,Goecke:2012qm}, the expansion
    of Fig.~\ref{fig:lbl_contributions} has been employed and consequently the exchanged mesons were considered off-shell.
    In this work we take a different perspective and adopt the viewpoint of the dispersive approach: that individual resonant
    contributions can be exactly evaluated for form factors taken on-shell, at the cost of needing to include all resonances
    also beyond single particle exchanges. Note that since this is a different expansion, any numbers that we present below
    do not supersede those presented in the past \cite{Fischer:2010iz,Goecke:2010if,Goecke:2012qm}, but represent new results.
    They will be compared to corresponding numbers from the fully data-driven dispersive approach. The difference of our
    approach to the fully dispersive one is in the evaluation of the various (space-like) EMFFs and TFFs
    needed to evaluate the different contributions: whereas in the dispersive approach these form factors are extracted
    from (mostly time-like) experimental data using analytic continuation, we calculate them directly from the underlying
    quark-gluon interaction at space-like momenta. Thus the two approaches nicely complement each other.

\subsection{Pion box contributions}\label{sec:pionbox}

\begin{figure}[!b]
      \begin{center}
      \includegraphics[width=0.77\columnwidth]{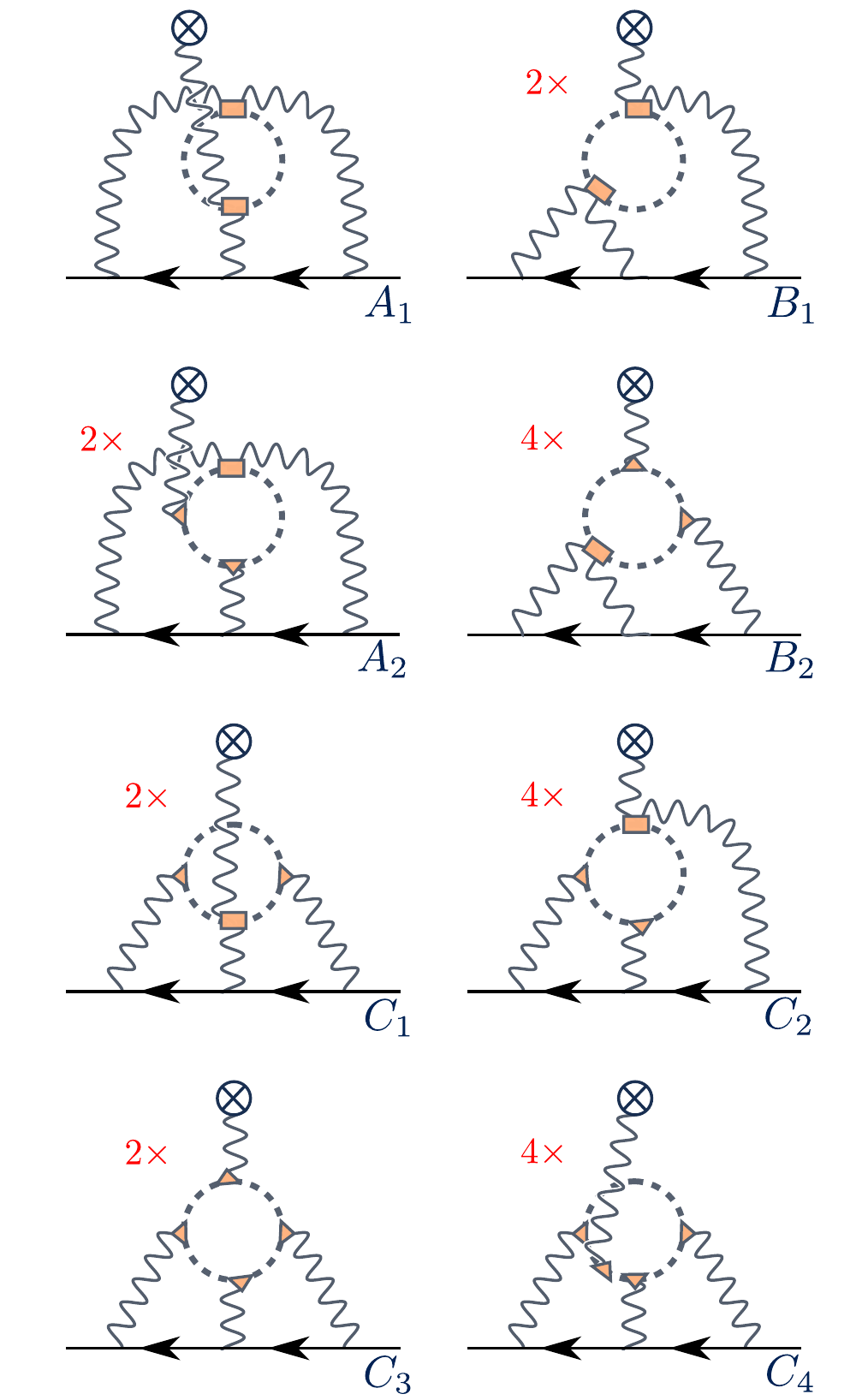}
      \caption{Pion box contributions to the muon g-2 in the framework of scalar QED.\label{fig:hlbl_pion_loop} }
      \end{center}
      \end{figure}

    It has been demonstrated in Ref.~\cite{Colangelo:2015ama} that the pion-box topology of the dispersive
    approach associated with pion-loop contributions coincides with the one-loop amplitude of scalar QED when coupled
    with pion form factors (FsQED). The basic observation is that the pion EMFFs $F_\pi(q_i^2)$ only depend
    on the momenta $q_1$, $q_2$ and $q_3$ of the three internal photons and therefore do not affect the integration of
    the pion loop, which then reduces to the corresponding one of scalar QED. Such one-loop contributions to scalar QED,
    both with and without pion EMFFs, have similarly been considered in Ref.~\cite{Kinoshita:1984it}. We follow
    the procedure detailed therein, which requires the evaluation of the six classes of diagrams shown in Fig.~\ref{fig:hlbl_pion_loop}.

    \begin{figure*}[!t]
    \begin{center}
    \includegraphics[width=0.7\textwidth]{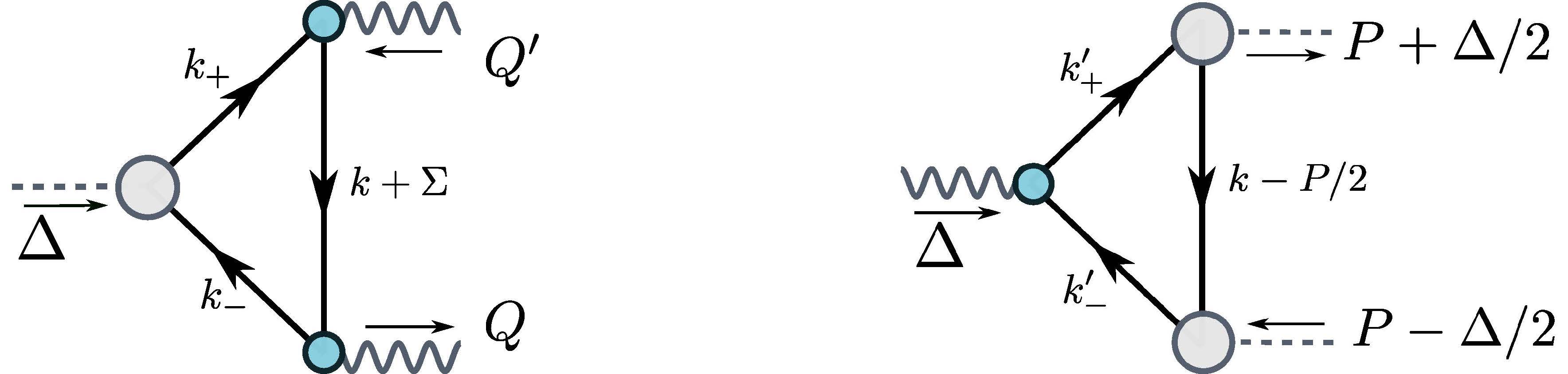}
    \caption{\textit{Left:} Meson transition form factor in rainbow-ladder truncation. The non-perturbative ingredients are the meson
    Bethe-Salpeter amplitude $\Gamma_M$ (gray circle), the dressed quark propagators (straight lines)
    and the dressed quark-photon vertices $\Gamma_\nu$ (blue circles).
    The internal momenta are $k_\pm = k \pm \Delta/2$, where
    $\Delta$ is the total momentum and $\Sigma = (Q+Q')/2$ the average momentum of the photons. 
    The average momenta entering the vertices are $r_\pm = k + \Sigma/2 \pm \Delta/4$.
    \textit{Right:} Analogous diagram for the pion form factor with internal momentum routing $k'_\pm = k + P/2 \pm \Delta/2$
    and average momenta $r_\pm' = k \pm \Delta/4$ in the Bethe-Salpeter amplitudes.
    \label{fig.pigg:diagram} }
    \end{center}
    \end{figure*}

\section{Electromagnetic and transition form factors}\label{sec:meson_form_factors}

In the following we briefly outline the various steps needed to calculate the pseudoscalar-meson EMFFs and TFFs
in the functional DSE approach. Details can be found in
\cite{Maris:1999bh,Maris:2002mz,Goecke:2010if,Goecke:2012qm,Raya:2015gva,Raya:2016yuj,Eichmann:2017wil}
and the review articles \cite{Maris:2003vk,Maris:2005tt,Eichmann:2016yit}. Diagrammatically, these
form factors are calculated as shown in Fig.~\ref{fig.pigg:diagram}.

The pseudoscalar TFF $F_{M\gamma\gamma}(Q^2,{Q'}^2)$ can be extracted from the transition matrix element via
\begin{equation}\label{pigg-current}
\begin{split}
	\mathcal{F}^{\mu\nu}_M(Q,Q')  &= e^2\,F_{M\gamma\gamma}(Q^2,{Q'}^2)\,\varepsilon^{\mu\nu\alpha\beta}  {Q'}^\alpha Q^\beta  \\
	                       &= 2e^2\, \text{Tr} \int_k \,  S_\q(k_+)\,\Gamma_M(k,\Delta)\,S_\q(k_-) \\
                           & \quad \times \Gamma^\mu(r_-,-Q)\,S_\q(k+\Sigma)\,\Gamma^\nu(r_+,Q')\,,
\end{split}
\end{equation}
where $Q$ and $Q'$ are the photon momenta and $e^2=4\pi \alpha_\text{em}$ is the squared electromagnetic charge.
The triangle diagram contains the dressed quark propagator $S_\q$,
the Bethe-Salpeter amplitude $\Gamma_M $ of the pseudoscalar meson $M$ and the dressed quark-photon vertex $\Gamma^\mu$ as shown in
the left panel of Fig.~\ref{fig.pigg:diagram}.
$F_{M\gamma\gamma}(Q^2,{Q'}^2)$ is dimensionful; in the chiral limit $F_{\pi\gamma\gamma}(0,0) = 1/(4\pi^2 f_\pi^0)$
due to the Abelian anomaly, where $f_\pi^0$ is the pion's electroweak decay constant in the chiral limit.

Similarly, the pion EMFF $F_\pi(\Delta^2)$ is extracted from the on-shell $\gamma \pi \pi$ current
in the right panel of Fig.~\ref{fig.pigg:diagram} via
\begin{equation}\label{onshell-current}
\begin{split}
	J^\mu(P,\Delta) &= 2 P^\mu F_\pi(\Delta^2) \\
               &= \Tr\!\!\int_k  S_\q(k_+')\,\Gamma^{\mu}(k,\Delta) \,S_\q(k_-') \\
               & \;\; \times \Gamma_{\pi}(r_-',P_-) \,S_\q(k-P/2)\, \Gamma_{\pi}(r_+',-P_+) \,,
\end{split}
\end{equation}
where $\Delta$ is the photon momentum, $P$ the average pion momentum and $P_\pm = P \pm \Delta/2$.

The necessary input to both Eqs.~\eqref{pigg-current} and~\eqref{onshell-current} is determined from a combination
of DSEs and BSEs. The Bethe-Salpeter amplitude of a pseudoscalar meson
and the quark-photon vertex satisfy (in-)homogeneous BSEs
\begin{align}
   [\Gamma_M(p,P)]_{\alpha\beta} &= \int_q  [ \mathbf{K}(p,q,P)]_{\alpha\gamma;\delta\beta}   \nonumber \\
                                 & \quad \times [S_\q(q_+)\, \Gamma_M(q,P)\,S_\q(q_-)]_{\gamma\delta}\,,  \label{mesonBSE} \\
   [\Gamma^{\mu}(p,P)]_{\alpha\beta} &= Z_2\,i\gamma^\mu_{\alpha\beta} + \int_q  [ \mathbf{K}(p,q,P)]_{\alpha\gamma;\delta\beta} \nonumber  \\
                                 & \quad \times [S_\q(q_+)\, \Gamma^\mu(q,P)\,S_\q(q_-)]_{\gamma\delta}\,,  \label{qqgamma}
\end{align}
where $\mathbf{K}$ is the Bethe-Salpeter kernel, $Z_2$ the quark renormalization constant, and in both equations $q_\pm = q \pm P/2$.
The quark propagator $S_\q$ is given by its DSE,
\begin{equation}
\begin{split}
 S^{-1}_\q(p) &= Z_2\,(i\slashed{p} + Z_m m_q) \\
              &- Z_{1f} \,g^2 \,C_F\int_q i\gamma^\mu  \, S_\q(q) \, \Gamma^\nu_\text{qg}(q,p)\,D^{\mu \nu}(k) \,,
\end{split}
\end{equation}
where $m_q$ is the current-quark mass, $k=q-p$, $C_F=4/3$,
$D^{\mu \nu}$ is the dressed gluon propagator, $\Gamma^\nu_\text{qg}$ the dressed quark-gluon vertex
and $Z_2$, $Z_m$ and $Z_{1f}$ are renormalization constants.
The gluon propagator and quark-gluon vertex satisfy their own DSEs which include further $n$-point functions, so that in all practical applications
the tower of DSEs needs to be truncated.

In the following we work in Landau gauge and use the rainbow-ladder truncation,
which together with more advanced schemes has been reviewed recently in Ref.~\cite{Eichmann:2016yit}. To this end one
defines an effective running coupling $\alpha(k^2)$ that incorporates dressing effects of the gluon propagator and
the quark-gluon vertex. In the quark DSE this entails
\begin{equation}
   Z_{1f}\,g^2\,\Gamma^\nu_\text{qg}(q,p)\,D^{\mu \nu}(k) \; \to \; Z_2^2\,\frac{4\pi\alpha(k^2)}{k^2}\,T^{\mu\nu}_k\,i\gamma^\nu
\end{equation}
with transverse projector $T^{\mu \nu}_k = \delta^{\mu\nu} - k^\mu k^\nu/k^2$.
The kernel $\mathbf{K}$ in the BSEs (\ref{mesonBSE}--\ref{qqgamma}) is uniquely related to the
quark-self energy by an axialvector Ward-Takahashi identity. In rainbow-ladder truncation it is given by
\begin{equation}
   [ \mathbf{K}(p,q,P)]_{\alpha\gamma;\delta\beta} \; \to \; Z_2^2\,\frac{4\pi\alpha(k^2)}{k^2}\, i\gamma^\mu_{\alpha\gamma}\,T^{\mu\nu}_k\, i\gamma^\nu_{\delta\beta}\,.
\end{equation}
This construction satisfies chiral constraints such as the Gell-Mann-Oakes-Renner relation and ensures the
(pseudo-)Goldstone boson nature of the pion. Once we have specified the explicit shape of the effective
interaction $\alpha(k^2)$, all elements of the calculation of the form factors
follow and there is no room for any additional adjustments.

Similarly to our previous work on the pion TFF~\cite{Eichmann:2017wil} we use the
Maris-Tandy model for the effective coupling $\alpha(k^2)$, Eq.~(10) of Ref.~\cite{Maris:1999nt},
with parameters $\Lambda=0.74$ GeV and $\eta = 1.85 \pm 0.2$ (the parameters $\omega$ and $D$ therein
are related to the above via $\omega D = \Lambda^3$ and $\omega=\Lambda/\eta$). The scale $\Lambda$ is
fixed via experimental input; we use the pion decay constant for this purpose. The variation of $\eta$
then changes the shape of the quark-gluon interaction at small momenta, cf.~Fig. 3.13 in Ref.~\cite{Eichmann:2016yit},
and we use it in the following as a rough estimate of the truncation error. We work in the isospin symmetric
limit of equal up/down quark masses. With a current light quark
mass of $m_q=3.57$ MeV at a renormalization point $\mu=19$~GeV we obtain a pion mass and pion decay constant of
$m_{\pi^0} = 135.0(2)$ MeV and $f_{\pi^0} = 92.4(2)$ MeV. With the strange-quark mass fixed at $m_s=85$ MeV
we obtain a kaon mass $m_{K} = 495.0(5)$ MeV.

The resulting dynamical mass function $M(p^2)$ for the dressed quark propagator has been discussed
around Fig.~11 in Ref.~\cite{Goecke:2010if}. The different dynamics due to the larger strange-quark mass
has potential consequences for the TFFs of the pseudoscalar $\eta$ and $\eta'$ mesons, which
will be discussed in the results section below.

To address the TFFs also for mesons with strangeness content, we need to consider
the effects of $\eta-\eta'$ mixing. To this end we start from the ideally mixed states
with flavor content $\pi^0 \sim (u\bar{u}-d\bar{d})/\sqrt{2}$, $\eta_n \sim (u\bar{u}+d\bar{d})/\sqrt{2}$ and $\eta_s \sim s\bar{s}$.
We denote their TFFs by $F_{\pi\gamma\gamma}$, $F_{n\gamma\gamma}$ and $F_{s\gamma\gamma}$ and their decay constants by $f_\pi$, $f_n$ and $f_s$,
respectively.
To account for the different flavor traces in the triangle diagram of Fig.~\ref{fig.pigg:diagram},
we define
\begin{equation}
   \left\{ F_{\pi\gamma\gamma}, \, F_{n\gamma\gamma}, \, F_{s\gamma\gamma} \right\} = \frac{1}{4\pi^2 f_\pi}
   \left\{ \F_\pi, \, \tilde{c}_n\,\F_n, \, \tilde{c}_s\,\F_s\right\}
\end{equation}
with $\tilde{c}_n = c_n\,f_\pi/f_n$, $\tilde{c}_s = c_s\,f_\pi/f_s$ and
\begin{equation}
   c_n = \frac{q_u^2+q_d^2}{q_u^2-q_d^2} = \frac{5}{3}\,, \quad
   c_s = \frac{\sqrt{2}\,q_s^2}{q_u^2-q_d^2} = \frac{\sqrt{2}}{3}\,,
\end{equation}
so that the dimensionless TFFs $\F_{\pi,n,s}(Q^2,{Q'}^2)$ mainly differ by the dynamics of the valence quarks.
In the DSE rainbow-ladder calculation they are continuously connected by changing the current-quark mass, which
yields $\F(0,0) = 1$ in the chiral limit, $\F_{\pi,n}(0,0) = 0.996$ at the physical $u/d$ mass and $\F_s(0,0) = 0.890$
at the strange-quark mass.

To proceed, we employ the two-angle mixing scheme in the quark flavour basis
\cite{Schechter:1992iz,Feldmann:1998sh,Feldmann:1998vh,Feldmann:1999uf}. In this scheme the physical $\eta$ and
$\eta'$ states are expressed in terms of the ideally mixed states above via
\begin{align}\label{mixing}
\left(\begin{array}{c} \eta   \\ \eta'  \\ \end{array} \right) = U(\phi)
\left(\begin{array}{c} \eta_n \\ \eta_s \\ \end{array} \right) ,
\end{align}
where
\begin{align}
U(\phi) =
\left(\begin{array}{rr}
      \cos \phi    & -\sin \phi    \\
      \sin \phi    &  \cos \phi  \\ \end{array} \right).
\end{align}

The corresponding decay constants of the $n$ and $s$ components of the physical states
follow this pattern,
\begin{align}\renewcommand{\arraystretch}{1.2}
\left(\begin{array}{cc}
      f_\eta^n    & f_\eta^s    \\
      f_{\eta'}^n & f_{\eta'}^s \\ \end{array} \right) = U(\phi)
      \left(\begin{array}{cc}
            f_n    & 0    \\
            0      & f_s \\ \end{array} \right)   .
\end{align}
In terms of flavour singlet and octet contributions to the decay constants of the physical mesons
the mixing pattern results in
\begin{align}\renewcommand{\arraystretch}{1.2}
\left(\begin{array}{cc}
      f_\eta^8    & f_\eta^1    \\
      f_{\eta'}^8 & f_{\eta'}^1 \\ \end{array} \right) =
      \left(\begin{array}{rr}
            \cos \theta_8    & -\sin \theta_1    \\[1mm]
            \sin \theta_8    &  \cos \theta_1  \\ \end{array} \right)
      \left(\begin{array}{cc}
            f_8    & 0    \\
            0      & f_1 \\ \end{array} \right)
\end{align}
with two different angles $\theta_1$ and $\theta_8$. The different quantities are related to each other via
\cite{Feldmann:1998sh}
\begin{equation}
\begin{split}
f_8 &= \sqrt{(f_n^2 + 2f_s^2)/3}\,, \quad \theta_8 = \phi-\arctan\frac{\sqrt{2} f_s}{f_n}\,, \\
f_1 &= \sqrt{(2f_n^2 +f_s^2)/3}\,, \quad \theta_1 = \phi-\arctan\frac{\sqrt{2} f_n}{f_s}\,.
\end{split}
\end{equation}

The explicit values for $f_n$, $f_s$ and the angle $\phi$ have been determined in a number of works,
see e.g. \cite{Escribano:2015yup} for an overview. For our calculations below we will use \cite{Feldmann:1998vh}
\begin{align}\label{mixingparameters}
\frac{f_n}{f_\pi} = 1.07 (2)\,, \quad \frac{f_s}{f_\pi} = 1.34 (6)\,, \quad \phi = 39.3^\circ (1.0^\circ)\,.
\end{align}
Using the chiral anomaly predictions with $\F_n(0,0) = \F_s(0,0) = 1$ and assuming that the mixing in Eq.~(\ref{mixing})
is momentum independent one finds relations for the mixing of the TFFs in the chiral limit (see e.g. \cite{Escribano:2015nra}).
These are then generalised to physical quark masses and lead to
\begin{align}\label{mixing3} \renewcommand{\arraystretch}{1.1}
\left(\begin{array}{c} F_{\eta \gamma \gamma}(Q^2,{Q'}^2)   \\ F_{\eta' \gamma \gamma}(Q^2,{Q'}^2)  \end{array} \right) = U(\phi)
\left(\begin{array}{c} F_{n \gamma \gamma}(Q^2,{Q'}^2) \\ F_{s \gamma \gamma}(Q^2,{Q'}^2)  \end{array} \right) ,
\end{align}
which we will use below to determine our results for the $\eta$ and $\eta'$
TFFs.  Under the approximation $\F_\pi = \F_n = \F_s$ Eq.~\eqref{mixing3} simplifies to
\begin{align}\label{mixing4} \renewcommand{\arraystretch}{1.1}
\left(\begin{array}{c} F_{\eta \gamma \gamma}(Q^2,{Q'}^2)   \\ F_{\eta' \gamma \gamma}(Q^2,{Q'}^2)  \end{array} \right) =
           F_{\pi \gamma \gamma}(Q^2,{Q'}^2) \,U(\phi)
\left(\begin{array}{c} \tilde{c}_n \\ \tilde{c}_s  \end{array} \right)
\end{align}
which was used in
\cite{Roig:2014uja} (see also \cite{Guevara:2018rhj} for a recent update) to determine the $\eta$ and $\eta'$
TFFs from $F_{\pi \gamma \gamma}$. In the results section
below we will compare the TFFs from (\ref{mixing3}) and (\ref{mixing4}) in order to assess the relevance of the
different dynamics of the strange quark.

\begin{figure*}[!t]
\begin{center}
\includegraphics[width=0.5\textwidth]{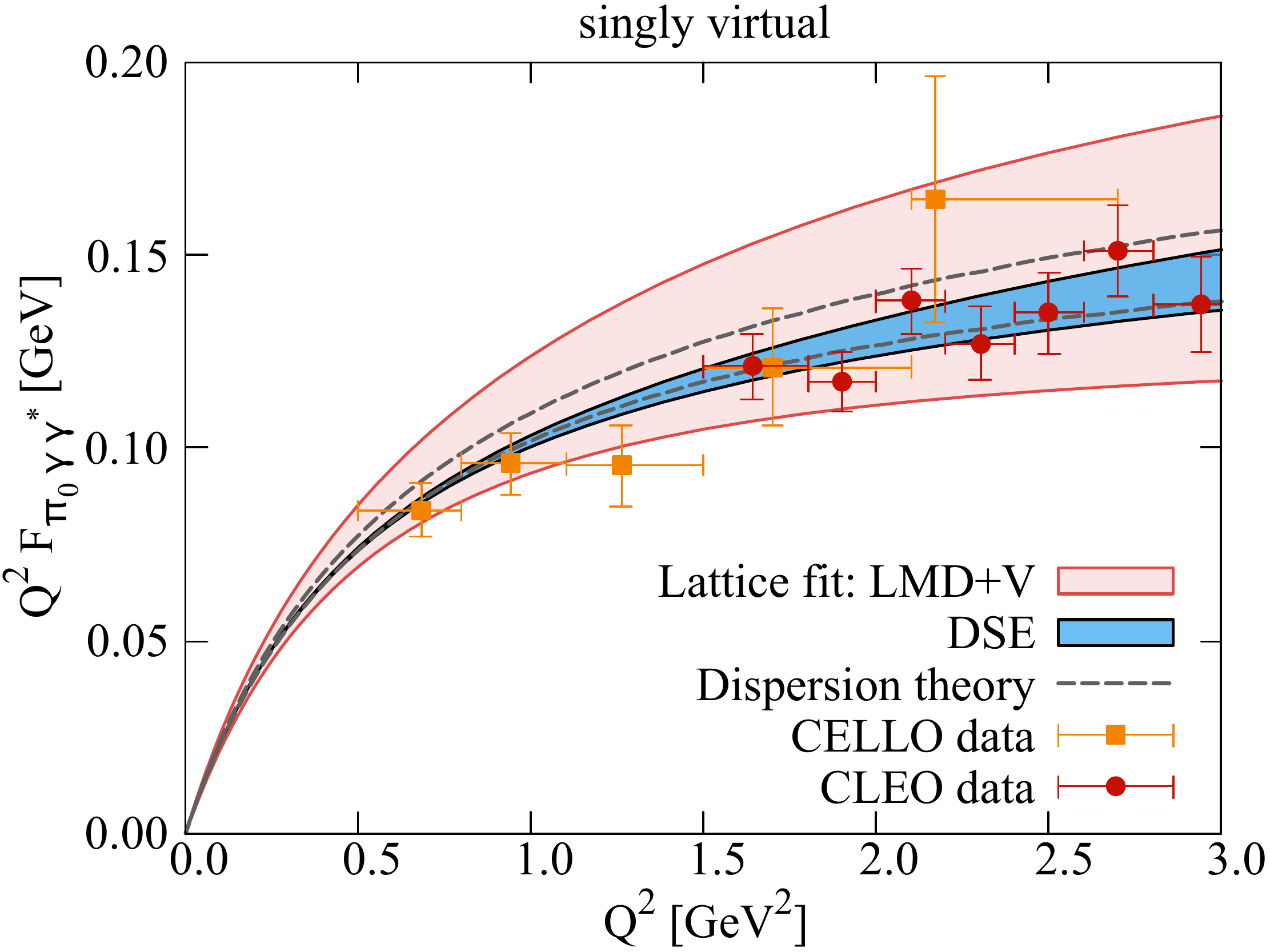}\hfill
\includegraphics[width=0.5\textwidth]{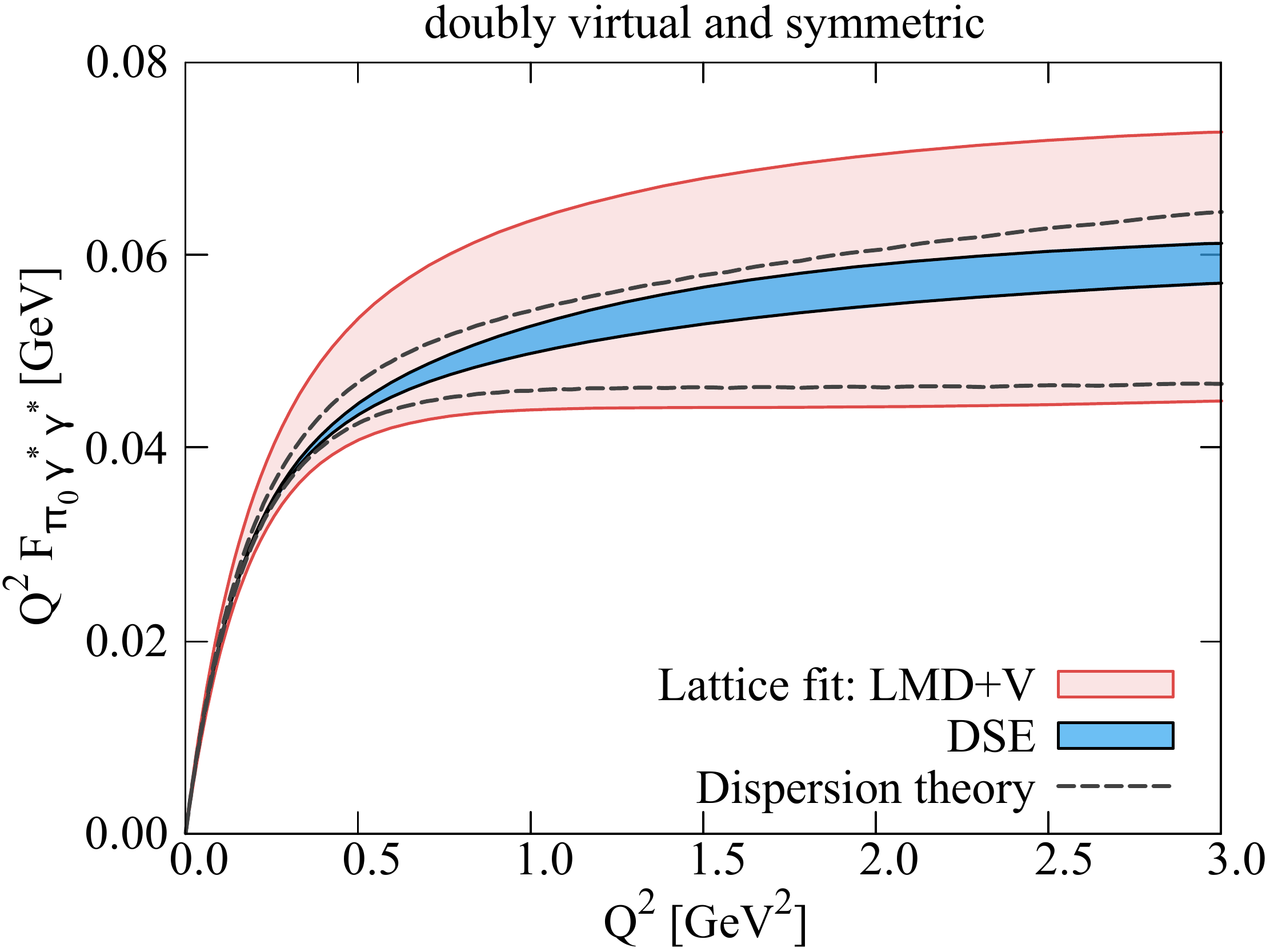}
\caption{The transition form factor $F_{\pi_0 \gamma \gamma}$ in the singly-virtual limit ($Q'^2=0$) and
the doubly virtual and symmetric limit ($Q'^2=Q^2$).
Compared are the results from the DSE approach \cite{Eichmann:2017wil}, lattice gauge theory \cite{Gerardin:2016cqj}
and dispersion theory \cite{Hoferichter:2018dmo,Hoferichter:2018kwz} together with the experimental results from
CELLO \cite{Behrend:1990sr} and CLEO \cite{Gronberg:1997fj} . \label{TFF:diagram} }
\end{center}
\begin{center}
\includegraphics[width=0.5\textwidth]{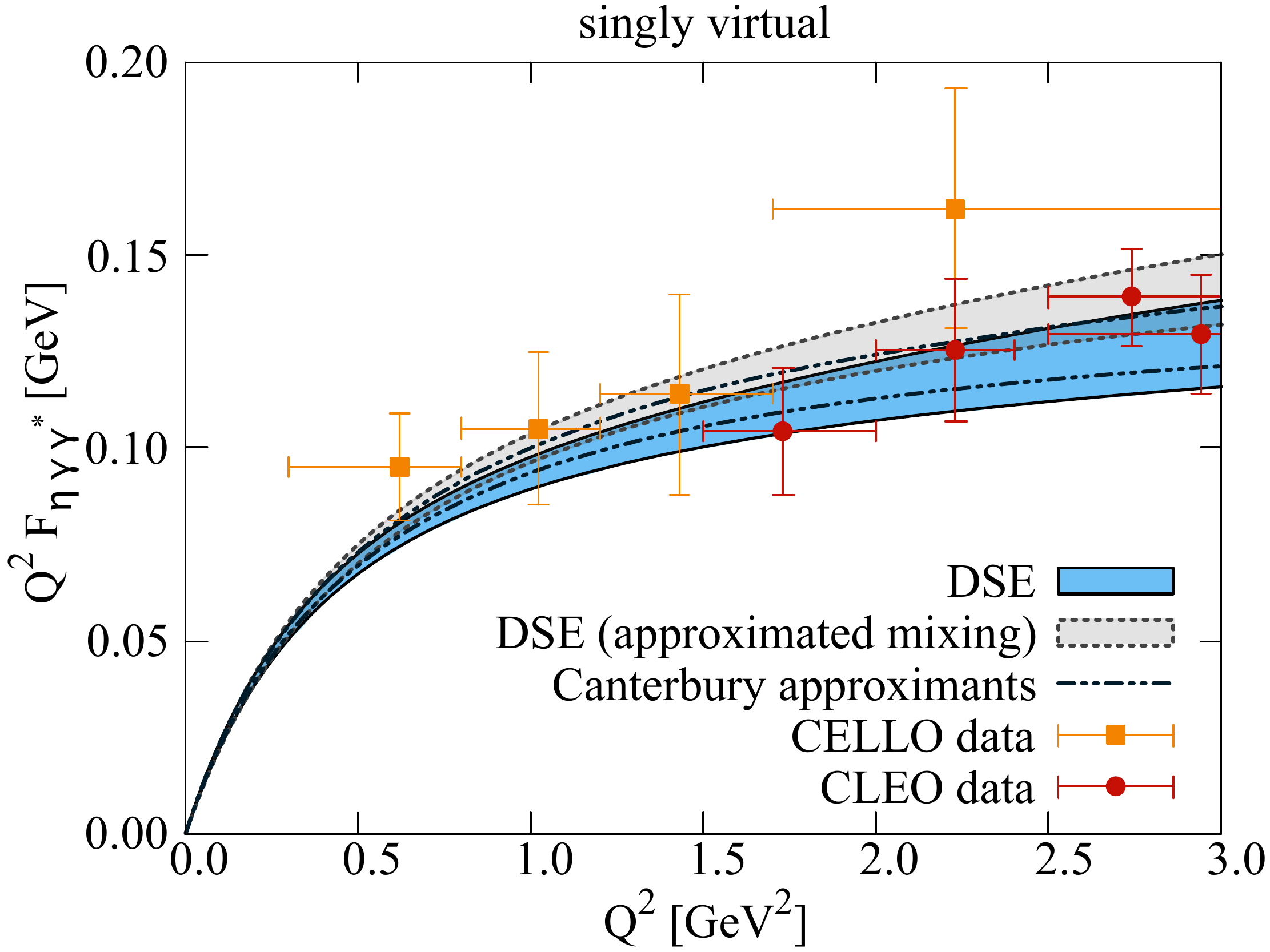}\hfill
\includegraphics[width=0.5\textwidth]{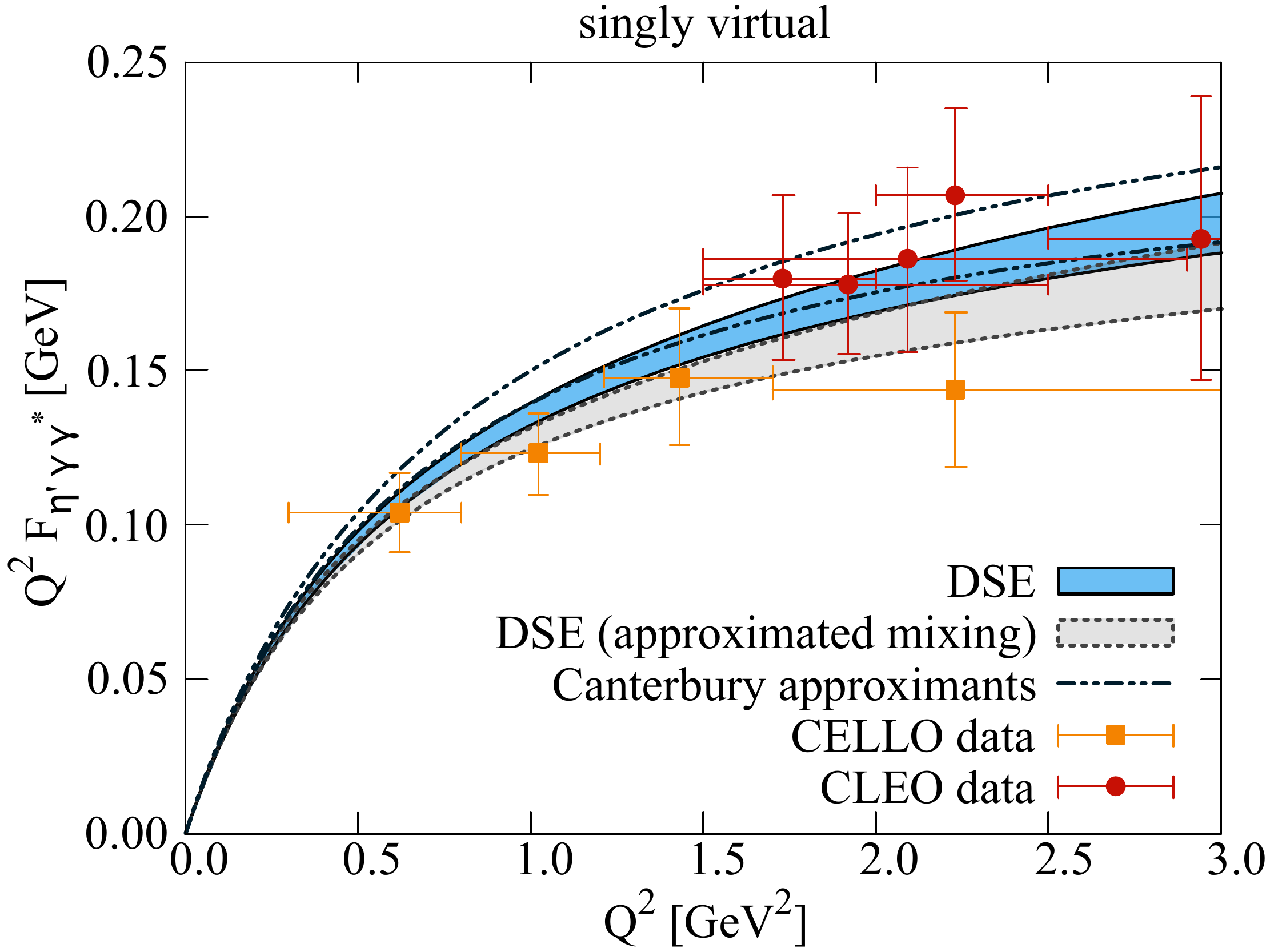}
\caption{The transition form factors $F_{\eta \gamma \gamma}$ and $F_{\eta' \gamma \gamma}$ in the
singly-virtual limit ($Q'^2=0$) compared to experimental results from CELLO \cite{Behrend:1990sr} and
CLEO \cite{Gronberg:1997fj} (results from different reactions averaged and error added in quadrature) and the results from a data based analysis using Canterbury approximants \cite{Masjuan:2017tvw}. \label{etaTFF:diagram} }
\end{center}
\end{figure*}
\begin{table*}[t!]
\begin{tabular}{c@{\;\;}| @{\;\;}c @{\;\;} | @{\;\;}c @{\;\;} | @{\;\;}c @{\;\;} | @{\;\;}c @{\;\;} | @{\;\;}c @{\;\;} | @{\;\;}c @{\;\;} | @{\;\;}c @{\;\;} | @{\;\;}c @{\;\;} | @{\;\;}c @{\;\;} | @{\;\;}c @{\;\;} | @{\;\;}c @{\;\;} | @{\;\;}c @{\;\;} | @{\;\;}c @{\;\;} | @{\;\;}c}
& $m_v$				& $a_0$	& $a_1$	& $a_2$	& $a_3$	& $b_1$	& $b_2$	& $b_3$	& $c_1$	& $c_2$	& $d_1$	& $d_2$	& $e_1$	& $e_2$ \\\hline	
$\F_{\pi}$	
& 0.77 \mbox{GeV}	& 0.996	& 0.735	& 1.214 & 1.547 & 0.089 & 0.133 &0.0002 & 0.384 & 0.430 & 2.010 & 0.024 & 1.540 & 0.00005 \\\hline
$\F_{s}$	
& 1.02 \mbox{GeV}	& 0.890	& 1.016	& 1.181 & 1.493 & 1.140 & 0.043 &0.00002& 0.418 & 0.489 & 2.220 & 0.101 & 1.540 & 0.00005
\end{tabular}
\caption{Parameters for the fit function Eq.~(\ref{fit}) for the pion and the strange-quark transition form factor.
\label{tab:param}}
\end{table*}

\section{Results}
\subsection{Pseudoscalar transition form factors}\label{sec:pseudoscalar_tff}

With all ingredients described in the previous section put together, numerical results for TFFs
in the functional approach have been discussed in a number of works,
see \cite{Maris:1999bh,Maris:2002mz,Goecke:2010if,Goecke:2012qm,Raya:2015gva,Raya:2016yuj,Eichmann:2017wil}
and references therein. It has been reported in \cite{Eichmann:2017wil} that the numerical data for the
pion TFF at space-like photon momenta $Q^2$ and $Q'^2$ can be accurately represented
by a suitable fit function. It turns out that the corresponding results for $F_{s \gamma \gamma}(Q,Q')$
can be reproduced with the same fit function but adapted parameters.
Abbreviating $w=(Q^2+Q'^2)/(2 m_v^2)$
and $z=(Q^2-Q'^2)/(Q^2+Q'^2)$, the momentum dependence of both TFFs is accurately described by
\begin{equation}
 \F_{\pi,s}(Q^2,{Q'}^2) = \frac{\mathcal{A}(w) + w(1-z^2)\,\mathcal{B}_1(w)\,(1+\mathcal{B}_2(w) z^2)} {(1+w)^2-w^2 z^2}
\end{equation}
with $\F_n = \F_\pi$.
The denominator represents the lowest-lying vector-meson pole 
corresponding to $Q^2=-m_v^2$
and ${Q'}^2 = -m_v^2$. The functions in the numerator ensure that the TFF asymptotically approaches a monopole
behaviour both in the symmetric (doubly-virtual) limit $z=0$ and the asymmetric (singly-virtual) limit $z=\pm 1$. They are given by
\begin{equation}\label{fit}
\begin{split}
  \mathcal{A}(w) &= \frac{a_0+\xi\,(a_1\, b_1\, w+a_2 \,b_2 \,w^2+a_3 \,b_3\,w^3)}{1+b_1 \,w+b_2 \,w^2+b_3 \,w^3}\,, \\
  \mathcal{B}_i(w) &= \frac{c_i \,e_i \,w^2}{1+d_i \,w + e_i \,w^2}\,.
\end{split}
\end{equation}
The parameter sets for the pion and $s\bar{s}$ TFFs are collected in Table~\ref{tab:param}.

This fit provides the input for our calculation of the pion pole contribution to $a_\mu$. The value $\xi = 1.0 \pm 0.1$
reflects our combined theoretical uncertainty for both fits from varying the parameter $\eta = 1.85 \pm 0.2$ in the
effective interaction as well as the uncertainty in the determination of the TFF away from the symmetric limit.

It is instructive to compare the pion TFF from the functional approach with the ones extracted from
dispersion theory \cite{Hoferichter:2018dmo,Hoferichter:2018kwz} and from lattice QCD \cite{Gerardin:2016cqj}.
This is shown for singly virtual asymmetric kinematics (i.e. $Q'^2=0$) and doubly virtual symmetric kinematics
in the two plots of Fig.~\ref{TFF:diagram}. In the momentum range displayed, which is most relevant
for $a_\mu$, all three approaches agree with each other and with the experimental data from the
CELLO \cite{Behrend:1990sr} and CLEO \cite{Gronberg:1997fj} collaborations within error bands. The result from the
functional DSE framework~\cite{Maris:2002mz, Eichmann:2017wil} moreover nicely agrees with the
dispersive result \cite{Hoferichter:2018dmo,Hoferichter:2018kwz} both in the zero momentum limit
and at larger momenta. Consequently, also the resulting slope parameter
\begin{align}
b_M &= m_M^2 \left.\frac{\partial}{\partial Q^2} \frac{F_{M \gamma \gamma}(0,Q^2)}{F_{M \gamma \gamma}(0,0)}\right|_{Q^2=0}
\end{align}
for $M=\pi_0$ from the DSE approach,
\begin{align}
b_\pi &= 31.10 \,(10) \times 10^{-3} \, ,
\end{align}
agrees well within error bars with the dispersive result $b_\pi=31.50 (90) \times 10^{-3}$
\cite{Hoferichter:2018dmo,Hoferichter:2018kwz}.

We wish to emphasize again that this agreement is not forced by any tuning of parameters. The DSE results
summarised here have been published already some time ago \cite{Maris:2002mz,Goecke:2010if,Eichmann:2017wil}.
Thus they were predictions, now confirmed by dispersive and lattice results. They also have been cross-checked
by determining rare decays of the $\pi^0$~\cite{Weil:2017knt}. As already discussed above,
the challenging part of the DSE calculation is a reliable determination of the total error budget.
The error quoted above and shown in the plot is a rough guess based on the variation of the one model parameter
$\eta$ in the effective interaction in the previously mentioned range. It is not a measure for the total truncation
error, which remains unaccounted for. Nevertheless it is satisfactory to see that the error bands agree with the
ones given by the dispersive approach for the singly virtual case. For the doubly virtual case the
estimated error of the DSE results is well within the error band of the dispersive results and it will be interesting to see whether
this remains so with further increasing precision of experimental data input for dispersion theory.

Our results for the $\eta$ and $\eta'$ TFFs in the singly virtual case are shown in
Fig.~\ref{etaTFF:diagram}. We compare results from the approximate mixing scheme~(\ref{mixing4}) with the
full scheme~(\ref{mixing3}) and experimental data from the CELLO~\cite{Behrend:1990sr} and CLEO~\cite{Gronberg:1997fj}
collaborations.
In addition, we show results from a data based analysis using Canterbury approximants \cite{Masjuan:2017tvw}.
The error bands of the DSE results are combined errors due to the variation of the parameter in the
interaction and the errors in the mixing parameters $\phi$, $f_n$ and $f_s$.
The effects of the dynamics of the strange quark are small in the low momentum regime and only
become noticeable for momenta larger than $Q^2 > 0.5 \,\mbox{GeV}^2$. At the largest momenta shown in the plot
the discrepancy between the approximated and full results is slightly larger than ten percent. The overall agreement
of our results with the experimental data and the results from the framework using Canterbury approximants is again excellent and we therefore feel confident to feed the TFFs in the corresponding meson exchange diagrams of LBL.

The slope parameters of our $\eta$ and $\eta'$ transition form factors are given by
\begin{align}
b_\eta    &= 0.51 \,(2)  \, , \qquad  b_{\eta'} = 1.57 \,(3) \, .
\end{align}
These are within the ballpark of other approaches including extractions from experiment,
see \cite{Escribano:2015yup,Escribano:2015nra, Masjuan:2017tvw} and references therein.

   \begin{figure*}[!t]
    \begin{center}
    \includegraphics[width=0.45\textwidth]{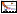}\hfill
    \includegraphics[width=0.55\textwidth]{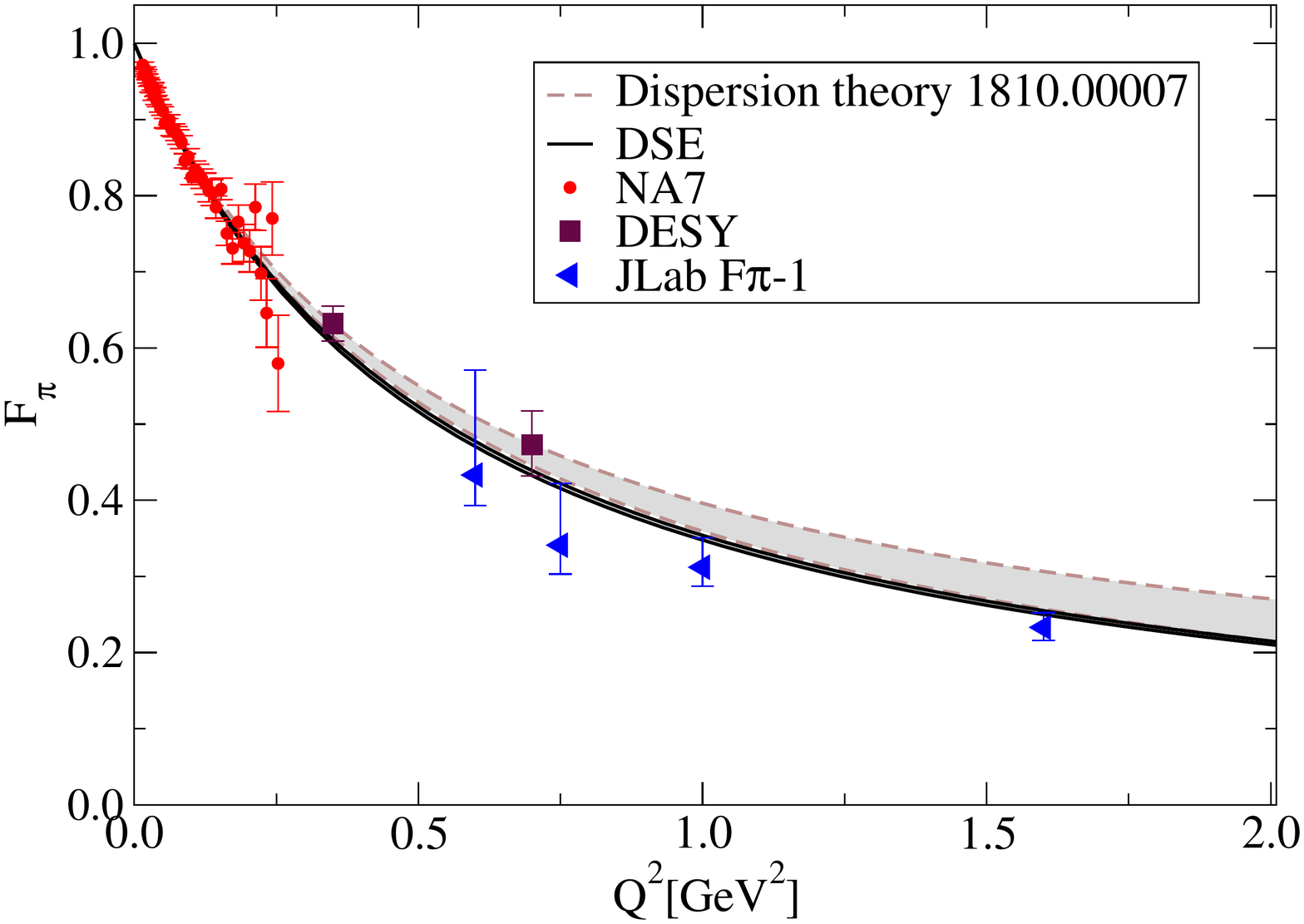}
    \caption{Left:The pion electromagnetic form factor as a function of the squared photon momentum.
    The experimental data have been extracted from Refs.~\cite{Ambrosino:2010bv,Fujikawa:2008ma,Amendolia:1986wj,Ackermann:1977rp,Brauel:1979zk,Volmer:2000ek,Horn:2006tm,Huber:2008id}. Right: Comparison with the dispersive results of \cite{Colangelo:2018mtw}. \label{pionFF} }
    \end{center}
    \end{figure*}

\subsection{Pion electromagnetic form factor}\label{sec:pion_ff}
    The EMFF of the pion $F_\pi(Q^2)$ in the rainbow-ladder truncation described above has already been determined
    in Refs.~\cite{Maris:1999bh,Krassnigg:2004if,Maris:2005tt}.
    For the purpose of the present work we have repeated this calculation
    including the variation of parameters in the effective interaction similar as for the TFFs.
    In Fig.~\ref{pionFF} we show the corresponding results compared with experimental data \cite{Ambrosino:2010bv,Fujikawa:2008ma,Amendolia:1986wj,Ackermann:1977rp,Brauel:1979zk,Volmer:2000ek,Horn:2006tm,Huber:2008id}.
    There is excellent agreement with the data in the spacelike region, which
    extends to the first $\rho$ pole in the timelike domain.
    In the domain $Q^2 \sim -0.3 \dots 4$ GeV$^2$, the numerical results
    are well described by a monopole ansatz
    \begin{align}\label{pifit}
    F_{\pi}(Q^2) = \frac{1}{1 + Q^2/L^2}
    \end{align}
    with $L = 0.735(5)$ GeV. In the range $Q^2 \sim 0 \dots 0.3$ GeV$^2$, our results furthermore agree very well
    with recent results from dispersion theory \cite{Colangelo:2018mtw,Ananthanarayan:2018nyx} with discrepancies
    smaller than the experimental error bars in this momentum region.

    One should emphasize that our result is the sum of two nontrivially competing contributions originating
    from different parts of the quark-photon vertex in the triangle diagram
    (right panel of Fig.~\ref{fig.pigg:diagram}). One is the Ball-Chiu vertex~\cite{Ball:1980ay}, which satisfies
    electromagnetic gauge invariance and is thus responsible for charge conservation $ F_{\pi}(0)=1$
    as well as the asymptotic limit $Q^2\to \infty$ where it becomes a bare vertex.
    The other is the transverse part which carries the dynamical information encoded in the
    solution of the inhomogeneous BSE~\eqref{qqgamma};
    it vanishes at the origin and contains the dynamically generated vector-meson poles.
    Fig.~\ref{pionFF} clearly shows that the effect of the transverse part is still sizeable in the mid-momentum region.
    For the TFFs discussed earlier the transverse part is even more important
    as it is necessary to reproduce the chiral-limit result for $F_{\pi\gamma\gamma}(0,0)$ stemming from
    the Abelian anomaly~\cite{Weil:2017knt}.

    We also note that the $\rho$ mesons in
    rainbow-ladder are stable bound states without widths and thus produce poles on the real axis in the form factor.
    Whereas our result $m_\rho = 0.744(1)$ GeV is in a similar range as the experimental $\rho(770)$ mass,
    the first radial excitation $\rho(1450)$ comes out far too light with $m_\rho' = 0.99(2)$ GeV.
    Because the poles are monopoles,
    $F_{\pi}$ proceeds from $-\infty$ at the first pole to $+\infty$ at the second,
    thus passing through zero in between; in the absolute value $|F_{\pi}|$ shown in Fig.~\ref{pionFF} this creates
    a zero followed by a pole. Adding widths by means of more sophisticated truncations would shift the poles
    into the complex plane \cite{Williams:2018adr,Miramontes:2018omq,Miramontes:2019mco}, but hardly affect
    the form factor in the space-like momentum region \cite{Miramontes:2019mco}. This is important for the
    stability of our results in the momentum range relevant for the evaluation of the pion box contribution
    to $a_\mu$.     

    The calculation of the EMFF beyond the $Q^2$ range displayed in Fig.~\ref{pionFF}
    faces the same obstacles as other matrix elements, namely the singularities of the
    $n$-point functions in the integrand which eventually cross the integration path.
    Above and below a certain $Q^2$ value these must be taken into account to obtain the correct result.
    For the pion TFF at large $Q^2$ we have circumvented the problem
    through interpolation between off-shell kinematics and the first $\rho$ pole~\cite{Eichmann:2017wil}, but
    the procedure is not directly applicable to the EMFF due to the different structure of the matrix element.
    A DSE-based determination of the pion EMFF at large $Q^2$ can be found in Ref.~\cite{Chang:2013nia}.

    In the right diagram of Fig.~\ref{pionFF} we also compare our results with the dispersive ones of 
    Ref.\cite{Colangelo:2018mtw}, which are an update of the results discussed in \cite{Colangelo:2017fiz}.
    For all $Q^2$ shown in the plot, the DSE results are slightly lower than the ones obtained 
    from dispersion theory. The large momentum behaviour of our result is in excellent agreement
    with the JLab-data. For the calculation of the pion box contribution to $a_\mu$ we have used 
    the fit, Eq.~(\ref{pifit}), in the entire momentum range tested by the diagram.

\subsection{Contributions to the anomalous magnetic moment of the muon}\label{sec:Results}
We finally proceed to discuss the pseudoscalar pole contributions and the pion box to the anomalous magnetic moment of
the muon $a_\mu$. As discussed in sections \ref{sec:mesonpole} and \ref{sec:pionbox}, these can be determined uniquely
once the corresponding TFFs and the pion EMFF are known. With the DSE results presented above
we obtain
\begin{align}\label{res:pionpole}
a_\mu^{\pi_0-\text{pole}} &= 62.6 \,(0.1)(1.3)  \times 10^{-11}\,.
\end{align}
The first error is due to the variation of the parameter $\eta$ in the effective interaction. From our results for the
TFF in Fig.~\ref{TFF:diagram} we find this variation to be very small for small momenta $Q^2$. Since
$a_\mu$ is dominated by contributions from small $Q^2$ of the order of the muon momentum this variation has almost no
effect. We have also added an additional second error of two percent for the numerical error accumulated from the calculation
of the quark functions in the DSEs, the Bethe-Salpeter amplitudes, the quark-photon vertex and the TFF.
The numerical error
in the actual calculation of $a_\mu$ from the TFFs is well under control and negligible. Since our TFF
is in very good agreement with the one determined by dispersion theory it is no surprise that
our value for $a_\mu^{\pi_0-\text{pole}}$ is as well: in \cite{Hoferichter:2018dmo}
$a_\mu^{\pi_0-\text{pole}} = 62.6^{+3.0}_{-2.5} \times 10^{-11}$ has been obtained. The framework of Ref.~\cite{Masjuan:2017tvw}
using Canterbury approximants resulted in $a_\mu^{\pi_0-\text{pole}} = 63.5 (1.2) (2.3) \times 10^{-11}$.
Further results from other groups can be found in \cite{Danilkin:2019mhd} and references therein.

Using the simple estimate for the $\eta$ and $\eta'$ TFFs discussed above, Eq.~(\ref{mixing4}), and the full result
including the dynamics of the strange quark, Eq.~(\ref{mixing3}), we also determined the corresponding contributions
to $a_\mu$ using the experimental values for their pole masses in the meson propagators. We obtain
\begin{align}
a_\mu^{\eta-\text{pole}} (\mbox{approx.})  &= 16.5 \,(0.3)(0.3)(1.1)  \times 10^{-11}\,, \\
a_\mu^{\eta'-\text{pole}} (\mbox{approx.}) &= 12.6 \,(0.3)(0.3)(0.6)  \times 10^{-11}\,,
\end{align}
and the full dynamical result
\begin{align}
a_\mu^{\eta-\text{pole}}  &= 15.8 \,(0.2)(0.3)(1.0)  \times 10^{-11}\,, \\
a_\mu^{\eta'-\text{pole}} &= 13.3 \,(0.4)(0.3)(0.6)  \times 10^{-11}\,.
\end{align}
Again, the first error stems from the variation of the model parameter and the second accounts for two percent numerical error.
The third error reflects the uncertainties in the mixing parameters (\ref{mixingparameters}). Our results are again in good
agreement with the treatment via Canterbury approximants of \cite{Masjuan:2017tvw}, where
$a_\mu^{\eta-\text{pole}}   = 16.2 \,(0.9)(0.9) \times 10^{-11}$ and
$a_\mu^{\eta'-\text{pole}}  = 14.5 \,(0.7)(1.7) \times 10^{-11}$ have been obtained (these are the averaged values of the
last two entries in their table II with error bars determined according to their prescription in the text).

It is very interesting to note that the sum of our $\eta$ and $\eta'$ pole contributions is identical for the simple
mixing scheme and the one including the full strange quark dynamics.
This is reasonable and well explained by comparing Eqs.~(\ref{mixing3}), (\ref{mixing4})
and the structure of the mixing matrix $U(\phi)$: With $\cos\phi$ and $\sin\phi$ of a similar magnitude and
$\tilde{c}_n \gg \tilde{c}_s$, the strange-quark effects are already suppressed in the individual $\eta$ and $\eta'$ TFFs,
whereas in the sum they almost cancel due to the opposite signs in $U(\phi)$. The different $\eta$ and $\eta'$
masses in the propagators which enter in $a_\mu^{\text{PS-pole}}$ do not change this behavior appreciably.

This induces a degree of stability into the total number
for the pseudoscalar pole contributions to $a_\mu$. We obtain
\begin{align}\label{res:pole}
a_\mu^{\text{PS-pole}}   &= 91.6 \,(1.9)  \times 10^{-11}\,,
\end{align}
where errors from different error sources are added in quadrature. 
This result agrees within error bars with the one obtained via Canterbury approximants
in Ref.~\cite{Masjuan:2017tvw}, which quotes $a_\mu^{\text{PS-pole}}= 94.3(5.3) \times 10^{-11}$. Again, further results from
other groups can be found in \cite{Danilkin:2019mhd} and references therein.

Naively comparing the pseudoscalar meson pole result~(\ref{res:pole}) with our earlier result for the pseudoscalar meson exchange
contributions $a_\mu^{\text{PS-exchange}} = 80.7 \times 10^{-11}$ \cite{Goecke:2010if} we find a difference of more than ten percent.
However, we wish to emphasise again that these contributions stem from different expansion schemes and should not be compared on
a one-to-one basis.

Finally, for the contribution of the pion box we obtain
\begin{align}\label{pion-box}
a_\mu^{\pi-\text{box}}   &= -16.3 \,(2)(4)  \times 10^{-11}\,,
\end{align}
where again the first error is due to the variation of the model parameter and the 
second accounts for our numerical error. Again, we obtain good agreement with the 
corresponding result from the dispersive approach within error bars:
the result given in Ref.~\cite{Colangelo:2017fiz} is 
$a_\mu^{\pi-\text{box}} = -15.9 \,(2)  \times 10^{-11}$. 

When comparing the errors of the two results one has to keep in mind that we solved the pion box
using the procedure of Ref.~\cite{Kinoshita:1984it} as outlined above in section 
\ref{sec:pionbox}. This involves the evaluation of a nine dimensional integral using
Monte-Carlo methods (we use the vegas routine from the Cuba library \cite{Hahn:2004fe}). 
On the other hand, 
the result in \cite{Colangelo:2017fiz} has been obtained after an elaborate analytical 
reformulation of the problem which allows to drastically reduce the numerical error.
In order to compare the two results it is thus useful to compare a control calculation:
in \cite{Colangelo:2017fiz} the authors give the result 
$a_\mu^{\pi-\text{box},\text{VMD}} = -16.4 \times 10^{-11}$ for a vector-meson dominance
type form factor. For the same VMD form factor we obtain 
$a_\mu^{\pi-\text{box},\text{VMD}} = -17.1 (4) \times 10^{-11}$, i.e. we (almost) 
agree within error bars, however our central value is somewhat too large. 
Since we are using the same SOBOL quasirandom sequence in all calculations this indicates 
that also the central value of our result (\ref{pion-box}) will become smaller when 
more accurate methods are used. This then is in agreement with the observation that 
our form factor is slightly lower than the dispersive one which should lead to 
a smaller value for $a_\mu^{\pi-\text{box}}$. 

Since our calculation involves a systematic truncation error which is hard to quantify
we see no merit in going through the same procedure as the authors of \cite{Colangelo:2017fiz}
to beat down the numerical error. On the contrary, since a smaller numerical error (i.e.
higher precision) might mislead readers into believing that the accuracy of our result 
is better than the one of the dispersive result (which due to the truncation error 
is not the case) we refrain from doing so.

\section{Summary}

In this work we have presented a calculation of the pseudoscalar pole and pion box contributions to hadronic light-by-light scattering based on a functional
approach to QCD via Dyson-Schwinger and Bethe-Salpeter equations. We employed a rainbow-ladder truncation for the
quark-gluon interaction that has emerged over the years as an excellent practical tool to obtain comprehensive results in the
pseudoscalar meson sector \cite{Maris:2003vk,Maris:2005tt,Horn:2016rip,Eichmann:2016yit}. Our results for the pion transition
form factor and, consequently, the pion pole contribution to $a_\mu$ are in excellent agreement with the most recent dispersive
result of Ref.~\cite{Hoferichter:2018dmo,Hoferichter:2018kwz}. Based on this agreement we consider our results for the $\eta$
and $\eta'$ pole contributions as quantitatively meaningful predictions. This assessment is supported by the very good agreement
with the results of a framework using Canterbury approximants \cite{Masjuan:2017tvw}. Observing that dynamical effects due to
the presence of the strange quark in the $\eta$ and $\eta'$ mesons cancel out in the sum of the two contributions, our value
for the total contribution $a_\mu^{PS-\text{pole}}$ of pseudoscalar meson poles is even stronger, since it can be founded on
the pion transition form factor alone.

The contribution (\ref{res:pole}) is accepted to be the leading part of the dispersive expansion of hadronic light-by-light scattering. Further
contributions are expected from scalar and axialvector pole contributions, see e.g. \cite{Pauk:2014rfa,Knecht:2018sci} and
references therein. These are also accessible in the functional approach and work in this direction is well under way.

\vspace*{3mm}
{\bf Acknowledgments}\\
We are grateful to Johan Bijnens, Gilberto Colangelo, Simon Eidelman, Bai-Long Hoid, Andreas Krassnigg,
Bastian Kubis, Stefan Leupold, Pablo Sanchez-Puertas and Hartmut Wittig for discussions.
This work was supported by the Helmholtz International Center for FAIR within
the LOEWE program of the State of Hesse, by the Helmholtz centre GSI in Darmstadt, Germany
and by the FCT Investigator Grant IF/00898/2015.

\bibliography{paper}

\end{document}